\RequirePackage{fix-cm}
\documentclass[onecolumn, final]{svjour3}   
\AtBeginDocument{%
	\setlength{\oddsidemargin}{\dimexpr(\paperwidth-\textwidth)/2-1in}%
	\setlength{\evensidemargin}{\oddsidemargin}%
	\setlength{\topmargin}{%
		\dimexpr(\paperheight-\textheight)/2-\headheight-\headsep-1in}%
}

\usepackage[dvipsnames]{xcolor}
\usepackage{graphicx}
\usepackage{comment}
\usepackage[authoryear, round]{natbib}
\usepackage{etoolbox}
\usepackage{xcolor}
\definecolor{darkblue}{RGB}{34,1,134}

\usepackage{hyperref} 
\hypersetup{
    colorlinks=true,
    linkcolor=darkblue,
    citecolor=darkblue,
    urlcolor=darkblue,
    pdftitle={Decolonial AI},
}

\usepackage{breakurl}

\citetop{S Mohamed, MT Png, W Isaac, Decolonial AI: Decolonial Theory \\
	as Sociotechnical Foresight in Artificial Intelligence, \\
	\textit{Philosophy and Technology} (405), 2020.}
\papernum{405}
\doinum{DOI: 10.1007/s13347-020-00405-8}
\journalname{Philosophy and Technology}
\begin{document}

\title{Decolonial AI: Decolonial Theory as \\Sociotechnical Foresight in Artificial Intelligence}
\titlerunning{Decolonial AI}

\author{Shakir Mohamed* \and Marie-Therese Png* \and William Isaac*}
\authorrunning{Mohamed, Png, Isaac }

\date{}

\institute{
Shakir Mohamed \at DeepMind, London 
\email{shakir@deepmind.com}         
\and  
Marie-Therese Png \at University of Oxford 
\email{marie-therese.png@oii.ox.ac.uk}
\and
William Isaac \at DeepMind, London
\email{williamis@deepmind.com} 
\and 
* Equal contributions. \\
\textit{Submitted 16 January, Accepted 26 May 2020}\\
To cite: S. Mohamed, MT. Png and W. Isaac (2020) Decolonial AI: Decolonial Theory as Sociotechnical Foresight in Artificial Intelligence. \textit{Philosophy and Technology} (405).
}

\maketitle

\begin{abstract}
\vspace{-10mm}
This paper explores the important role of critical science, and in particular of post-colonial and decolonial theories, in understanding and shaping the ongoing advances in artificial intelligence. Artificial Intelligence (AI) is viewed as amongst the technological advances that will reshape modern societies and their relations. Whilst the design and deployment of systems that continually adapt holds the promise of far-reaching positive change, they simultaneously pose significant risks, especially to already vulnerable peoples. Values and power are central to this discussion. Decolonial theories use historical hindsight to explain patterns of power that shape our intellectual, political, economic, and social world.
By embedding a decolonial critical approach within its technical practice, AI communities can develop foresight and tactics that can better align research and technology development with established ethical principles, centring vulnerable peoples who continue to bear the brunt of negative impacts of innovation and scientific progress. 
We highlight problematic applications that are instances of coloniality, and using a decolonial lens, submit three tactics that can form a decolonial field of artificial intelligence: creating a critical technical practice of AI, seeking reverse tutelage and reverse pedagogies, and the renewal of affective and political communities.
The years ahead will usher in a wave of new scientific breakthroughs and technologies driven by AI research, making it incumbent upon AI communities to strengthen the social contract through ethical foresight and the multiplicity of intellectual perspectives available to us; ultimately supporting future technologies that enable greater well-being, with the goal of beneficence and justice for all. 

\end{abstract}

\section{How Values Shape Scientific Knowledge and Technology}
\label{sect:vst}

The ongoing advances in artificial intelligence (AI), and innovations in technology more generally, encompass ever-larger aspects of the cultural, economic and political life of modern society. We aim to capture this expanding role and impact of AI by widening the conceptual aperture with which AI is understood: dually viewing AI as both {object and subject}, i.e. viewing AI as technological artefacts and as systems of networks and institutions, respectively.

As an object, advances in AI research\footnote{This view of AI as object, and the of the term AI throughout, will be used as an umbrella term that includes the field of machine learning. Both machine learning and artificial intelligence are disciplines focused on the science and engineering of intelligent agents or computer programs \citep{russell2016artificial,boden2018artificial}. While the broader field of AI includes both symbolic (also known as classical AI or GOFAI) and connectionist (e.g., artificial neural networks) research, the field of machine learning can be defined by research on more tractable machine tasks leveraging techniques at the intersection of computer science and statistical inference \citep{mitchell2006discipline}.} over the last two decades---often attributed to a combination of increases in computational power, availability of large amounts of data, and advances in learning algorithms \citep{lecun2015deep}---has led to novel applications in a wide range of sectors, including transportation and healthcare, amongst others \citep{gerrish2018smart}. While these recent innovations have led to some societal benefits, they have also demonstrated their potential to be abused or misused in ways their designers could not have imagined \citep{oneil2016weapons}.  As a subject, AI has seen itself elevated from an obscure domain of computer science into technological artefacts embedded within and scrutinised by governments, industry and civil society. These stakeholders play a significant role in shaping the future direction and use of advanced technologies such as AI---whether through the establishment of regulatory and ethical frameworks or the promotion of specific algorithmic architectures\footnote{Specific applications of machine learning and AI largely centre on the use of learning \textit{algorithms} that manipulate and transform data into information suitable for the given task.}---that warrants consideration under a more expansive conceptualisation of the term AI.

As both object and subject, the aims and applications of AI have been brought into question. At the heart of these discussions are questions of values and the power relations in which these values are embedded. What values and norms should we aim to uphold when performing research or deployment of systems based on artificial intelligence? In what ways do failures to account for asymmetrical power dynamics undermine our ability to mitigate identified harms from AI? How do unacknowledged and unquestioned systems of values and power inhibit our ability to assess harms and failures in the future?

\subsection{The Evolution of Value and Power Paradigms}

The role that values play in the process of generating new knowledge is a perennial question, particularly in the sciences. Philosophers have debated the importance of \textit{epistemic values}, such as internal consistency, falsifiability, generalisability of a particular theory, and notions of scientific objectivity \citep{laudan1968theories, bueter2015irreducibility}. These values shape the veracity of scientific statements, aiming to establish broader ontological or causal claims about the nature of specific systems. Yet, science is a product not only of epistemic values, but also of \textit{contextual values} that reflect moral, societal or personal concerns in the application of scientific knowledge. There is strong consensus that non-epistemic values have a legitimate role in scientific reasoning, particularly in the choice of research projects and the application of scientific results \citep{elliott2014nonepistemic, douglass2007}. This role of contextual values also applies to the work of computing and technology \citep{nissenbaum2001computer, van2014can}---a recognition established in the broader field of Values in Technology \citep{friedman2013value, sengers2005reflective, disalvo2012adversarial}.

Due to repeated instances of unethical research practices within the scientific community---instances like the U.S. Public Health Service Syphilis Study at Tuskegee \citep{brandt1978racism}---concerned scientists, policy-makers and human rights advocates responded by formalising contextual values into ethical frameworks that reoriented power relations between researchers and impacted communities or persons. Efforts such as the Nuremberg Code \citep{nurembergcode}, the Helsinki declaration \citep{helsinki1964}, and the Belmont Principles \citep{belmontreport} collectively led to the establishment of three core ethical values or rights that should serve as a minimum standard for human subject research: respect-for-persons (individual autonomy), beneficence (research designed to maximise societal benefit and minimise individual harm), and justice (research risks must be distributed across society). These principles are viewed as a historical milestone for research ethics, although their violations continue to occur, e.g., the ongoing questions of unethical blood exports during the West Africa Ebola epidemic \citep{freudenthal2019ebola}.
These principles are also questioned and subject to many reappraisals, which have highlighted their failures in capturing a range of emerging or novel harms, or an insufficiency in capturing the lived realities of under-represented groups \citep{shore2006re,vitak2016beyond}. 

The limitations of these value principles become clearer as AI and other advanced technologies become enmeshed within high-stakes spheres of our society. 
Initial attempts to codify ethical guidelines for AI, e.g., the Asilomar principles \citep{asilomar2017}, focused on risks related to lethal autonomous weapons systems and AGI Safety. Though both are critical issues, these guidelines did not recognise that risks in peace and security are first felt by conflict zones in developing countries \citep{garcia2019militarization}, or engage in a disambiguation of social safety and technical safety.  
Moreover, they did not contend with the intersection of values and power, whose values are being represented, and the structural inequities that result in an unequal spread of benefits and risk within and across societies.

An example of this nexus between values, power, and AI is a recent study by \citet{obermeyer2019dissecting}, which revealed that a widely used prediction algorithm for selecting entry into healthcare programs was exhibiting racial bias against African-American patients. The tool was designed to identify patients suitable for enrolment into a ``high-risk care management" programme that provides access to enhanced medical resources and support. Unfortunately, large health systems in the US have emphasised contextual values to ``reduce overall costs for the healthcare system while increasing value''  \citep{american2018augmented} or ``value for money" \citep{nhs2019} when selecting potential vendors for algorithmic screening tools at the expense of other values such as addressing inequities in the health system. As a result, the deployed algorithm relied on the predictive utility of an individual's health expenses (defined as total healthcare expenditure) indirectly leading to the rejection of African-American patients at a higher rate relative to white patients, denying care to patients in need, and exacerbating structural inequities in the US healthcare system \citep{nelson2002unequal}. As this example shows,
the unique manner in which AI algorithms can quickly ingest, perpetuate, and legitimise forms of bias and harm represents a step change from previous technologies, warranting prompt reappraisal of these tools to ensure ethical and socially-beneficial use. 

An additional challenge is that AI can obscure asymmetrical power relations in ways that make it difficult for advocates and concerned developers to meaningfully address during development. As \citet{benjamin2019race} notes, ``whereas in a previous era, the intention to deepen racial inequities was more explicit, today coded inequity is perpetuated precisely because those who design and adopt such tools are not thinking carefully about systemic racism''. Some scholars such as \citet{floridi2018ai4people} have highlighted that technologies such as AI require an expansion of ethical frameworks, such as the Belmont Principles, to include explicability (explanation and transparency) or non-malfeasance (do no harm). \citet{whittlestone2019ethical} conversely argue for a move away from enumerating new value criteria, and instead highlight the need to engage more deeply with the tensions that arise between principles and their implementation in practice. Similarly, we argue that the field of AI would benefit from dynamic and robust \textit{foresight} tactics and methodologies grounded in the critical sciences to better identify limitations of a given technology and their prospective ethical and social harms. 

\subsection{Critical Science as a Sociotechnical Foresight Tool}

The critical science approach represents a loosely associated group of disciplines that seek to uncover the underlying cultural assumptions that dominate a field of study and the broader society. Scholarship in this domain \citep{winner1980artifacts, nissenbaum2001computer,greene2019better} aims not only to explain sociotechnical phenomena, but to also examine issues of values, culture and power at play between stakeholders and technological artefacts. We use a necessarily broad scope of critical science theories due to the expansive range of applications of AI, but seek to emphasise particularly the role of \textit{post-colonial and decolonial critical theories}. While decolonial studies begins from a platform of historical colonialism, it is deeply entangled with the critical theories of race, feminism, law, queerness, and science and technology studies \citep{d2020data, feenberg2017critical}.

The role of values and power as they relate to technology and data has been argued by a multitude of scholars who draw from the decolonial theories, such as \citet{ricaurte2019data}, \cite{milan2019big} and \citet{couldry2019costs}, as well as established research in post-colonial and decolonial computing, such as those by \citet{irani2010postcolonial}, \citet{dourish2012ubicomp} and \citet{ali2016brief}. Such critical perspectives are increasingly used to elucidate potential ethical and social ramifications of AI and technology generally, with much research now available that exposes concerns of bias and injustice in algorithmic systems \citep{propublica2016, benjamin2019race,buolamwini2018gender, lum2016predict, noble2018algorithms,eubanks2018automating}, exploitative or extractive data practices \citep{gray2019ghost}, and applications of AI that dispossess the identity and resources of vulnerable populations \citep{green2019good,keyes2018misgendering, hanna2019towards, stark2019facial}.

In this paper, our aim is to guide readers through a brief introduction to decolonial theory, and to demonstrate how this theoretical framework can serve as a powerful lens of ethical foresight. Technology foresight and foresight methodologies more broadly is a term used to classify efforts by researchers, policy-makers and industry practitioners to understand and anticipate how choices and actions made today can shape or create the future \citep{coates1985foresight}. For AI technologies, problematic outcomes in high-stakes domains such as healthcare or criminal justice have demonstrated a clear need for dynamic and robust ethical foresight methodologies. Such methodologies could enable stakeholders to better anticipate and surface blind-spots and limitations, expand the scope of AI's benefits and harms, and reveal the relations of power that underlie their deployment. This is needed in order to better align our research and technology development with established and emerging ethical principles and regulation, and to empower vulnerable peoples who, so often, bear the brunt of negative impacts of innovation and scientific progress. 
\section{Coloniality and Decolonial Theory}
\label{sect:decolonisation}

Decolonisation refers to the intellectual, political, economic and societal work
concerned with the restoration of land and life
following the end of historical colonial periods \citep{ashcroft2006post}.
Territorial appropriation, exploitation of the natural environment and of human
labour, and direct control of social structures are the characteristics of
historical colonialism. Colonialism's effects endure in the present, and when these colonial characteristics are identified with present-day activities, we speak of the more general concept of \textit{coloniality} \citep{quijano2000coloniality, mignolo2007introduction, maldonado2007coloniality}. This section is a brief review of coloniality, its view on systems of power, and its manifestation in the digital world.

Coloniality is what survives colonialism \citep{ndlovu2015decoloniality}.
Coloniality therefore seeks to explain the continuation of power dynamics between those advantaged and disadvantaged by ``the historical processes of dispossession, enslavement, appropriation and extraction […] central to the emergence of the modern world'' \citep{bhambra2018decolonizing}. 
Coloniality names the continuity of established patterns of power between coloniser and colonised---and the contemporary remnants of these relationships---and how that power shapes our understanding of culture, labour, intersubjectivity, and knowledge production; what \citet{quijano2000coloniality} refers to as the coloniality of power. For  \citet{quijano2000coloniality} the power of coloniality lies in its control over social structures in the four dimensions of authority, economy, gender and sexuality, and knowledge and subjectivity. Similarly, for \citet{maldonado2007coloniality}, coloniality is the reproduction of hierarchies of race, gender and geopolitics, which were invented or instrumentalised as tools of colonial control. For \citet{couldry2019costs}, who bring the coloniality of power into the digital present, it is modern data relations---the human relations that when captured as data enables them to become a commodity---that ``recreate a colonising form of power''.

Consequently, \textit{decolonisation} takes two roles. The first is a territorial decolonisation that is achieved by the dissolution of colonial relations.
The second, is a structural decolonisation, with which this paper is concerned, that seeks to undo colonial mechanisms of power, economics, language, culture, and thinking that shapes contemporary life: interrogating the provenance and legitimacy of dominant forms of knowledge, values, norms and assumptions.
Three views clarify this decolonial knowledge landscape. 
\begin{itemize}
\item A \textit{decentring view} of decolonisation seeks to reject an imitation of the West in all aspects of life, calling for the assertion of unique
identities and a re-centring of knowledge on approaches that restore global
histories and problems and solutions. For Ng\~{u}g\~{\i} wa Thiong'o, this means
replacing the English language as the unassailable medium of teaching and
discourse \citep{wa1992decolonising}. Discussions on decolonising the
curriculum or decolonising the university call for reappraisals of what is
considered the foundation of an intellectual discipline by emphasising and recognising the legitimacy of marginalised knowledge
\citep{jansen2019decolonisation, bhambra2018decolonizing}; calls to decolonise
science often invoke this view of decolonisation
\citep{harding2011postcolonial}. 

\item An \textit{additive-inclusive view} continues
to use existing knowledge, but in ways that recognises explicitly the value of
new and alternative approaches, and that supports environments in which new ways
of creating knowledge can genuinely flourish. This view is invoked by works that
criticise universalism in thinking, and instead advocate for localisation and
pluriversalism \citep{mignolo2012local, escobar2011sustainability}. 
\item An
\textit{engagement view} calls directly for more critical views of science. This view calls on us to
examine scientific practice from the margins, to place the needs of marginalised populations at the centre of the design and research process, and to ask where knowledge comes from---who is included and left out, in whose interest is science applied, who is
silenced, and what unacknowledged assumptions might be at play \citep{mcdowell201630}.
\end{itemize}

Decolonial theory provides us with several useful tools with which to qualify the nature of power imbalances or inequitable impacts that arise from advanced technologies like AI.
One such framework identifies metropoles---the centres of power---and their peripheries that hold relatively less power and contest the metropole's authority, participation and legitimacy in shaping everyday life \citep{champion2005metropole}. 
Dependency theory expands on this framework, by tying colonial histories to present day underdevelopment and continued economic imbalance between countries and regions, as well as tying resulting dependencies to historic metropole and periphery dynamics \citep{champion2005metropole}. Using the lens of metropole and periphery, we can identify contemporary practices in AI development partially as features of colonial continuities from states and governments. Similarly, today's technology corporations could be described as metropoles of technological power with civic society and consumers sitting at the periphery.

Metropole-periphery dichotomies are interpretive models that if not used carefully can reduce the reality of lived experiences to overly-simplified binaries of `West and the rest', `North and South', `powerful and oppressed' \citep{mcclintock1992angel, stoler2008epistemic, thrush2008american}, exposing some of the limitations of decolonial theory. In addition, grand historical meta-narratives of injustice, contending with the theoretical idea of `global', and `speaking for the oppressed' \citep{pappas2017limitations} are pitfalls that must be avoided. A needed balance can be found by incorporating other modes of decolonial thought, such as contrapuntal analysis \citep{said2012culture}, psychodynamic perspectives \citep{fanon2007wretched, nandy1989intimate}, economic analysis \citep{pollard2011postcolonial}, and historical and literary criticism \citep{james1993beyond, gopal2019insurgent}, amongst others. Because of the limitations of the theory, we believe it is important to incorporate the wider critical science view introduced in the previous section.

\section{Algorithmic Coloniality}
\label{sect:coloniality_sites}
By recognising the analogues of territorial and structural coloniality in the digital age, we 
propose the application of
decolonial theory to digital technologies such as AI. Digital spaces---created by the internet and the increasingly-networked systems and devices we use---form digital territories that, like physical spaces, have the propensity to become sites of extraction and exploitation, and thus the sites of digital-territorial coloniality.
The coloniality of power can be observed in digital structures in the form of
socio-cultural imaginations, knowledge systems, and ways of 
developing and using technology which are based on systems, institutions, and values that
persist from the past and remain unquestioned in the present. As such,
emerging technologies like AI are directly subject to coloniality, giving decolonial critical theories a powerful analytical role.

The emerging theories of data colonialism \citep{thatcher2016data, ricaurte2019data, couldry2019costs} and data capitalism \citep{zuboff2019age} recognise this nature of historic continuity and the role of data as the material resource that is exploited for economic expansion. \citet{ricaurte2019data} develops a theoretical model that analyses the coloniality of technological power through data, examining data-centric epistemologies as an expression of the coloniality of power, in how they impose ``ways of being, thinking, and feeling that leads to the expulsion of human beings from the social order, denies the existence of alternative worlds and epistemologies, and threatens life on Earth'' \citep{ricaurte2019data}. \citet{couldry2019data} further expand on the colonial continuities of extraction and exploitation of land, labour and relations through digital infrastructure. This larger area of technological coloniality is further developed by scholars in areas of intersectional data feminism \citep{d2020data}, critical race theory \citep{benjamin2019race},  decolonisation of technology \citep{awori2016decolonising}, new data epistemologies \citep{milan2016alternative}, and environmental sustainability and justice \citep{ropke2001new, ricaurte2019data, gallopin1992science}. 

We use the term \textit{algorithmic coloniality} to build upon data colonialism in the context of the interactions of algorithms across societies, which impact the allocation of resources, human socio-cultural and political behaviour, and extant discriminatory systems. We also begin to examine how coloniality features in algorithmic decision-making systems as they generate new labour markets, impact geopolitical power dynamics, and influence ethics discourse. 

In the following section, we introduce the language of decoloniality to the current discourse on fairness, accountability, and transparency in algorithmic systems, as well as introduce a taxonomy of decolonial foresight: institutionalised algorithmic oppression, algorithmic exploitation, and algorithmic dispossession. Within these forms of decolonial foresight, we present a range of use-cases that we identify as sites of coloniality: algorithmic decision systems, ghost work, beta-testing, national policies, and international social development. By sites of coloniality we mean cases that exhibit structural inequalities that can be contextualised historically as colonial continuities. These sites of coloniality help identify where empirical observation departs from the current theoretical frameworks of power in AI, which by-and-large are ahistorical. By using these sites to address the clash of theory and empiricism, we argue that discussions of power and inequality as related to AI cannot be ahistorical and are incomplete if they fail to recognise colonial continuities. 

\subsection{Algorithmic Oppression}

Algorithmic oppression extends the unjust subordination of one social group and the privileging of another---maintained by a ``complex network of social restrictions'' ranging from social norms, laws, institutional rules, implicit biases, and stereotypes \citep{taylor2016groups}---through automated, data-driven and predictive systems. The notion of algorithmic or automated forms of oppression has been studied by scholars such as \citet{noble2018algorithms} and \citet{eubanks2018automating}. The following examples will make initial connections between instances of algorithmic oppression across geographies, and identify the role of decolonial theory in this discourse.

\subsubsection*{Site 1: Algorithmic Decision Systems}

Predictive systems leveraging AI have led to the formation of new types of policing and surveillance, access to government services, and reshaped conceptions of identity and speech in the digital age. Such systems were developed with the ostensible aim of providing decision-support tools that are evidence-driven, unbiased and consistent. Yet, evidence of how these tools are deployed shows a reality that is often the opposite. Instead, these systems risk entrenching historical injustice and amplify social biases in the data used to develop them \citep{benjamin2019race}. 

Evidence of such instances are abundant. As an example, digital human rights concerns have been widely-raised: in Singapore’s application of facial recognition in CCTV through the Lamppost as a Platform initiative (LaaP) \citep{johnston2019comparison}, New Delhi's CMAPS predictive policing system \citep{marda2020data}, India’s Aadhaar identity system \citep{siddiqui2015aadhar}, the Kenyan Huduma Namba digital/biometric identity system \citep{nyawa2019big}, and the welfare interventions for M\={a}ori children by the New Zealand government \citep{vaithianathan2013children,gavighan2019government}.
The impact of predictive algorithms on people, from everyday citizens to the most vulnerable, highlights the need for diversified and contextualised approaches to issues of justice and fairness in automated systems. 

Algorithmic decision systems \citep{isaac2017hope} are increasingly common within the US criminal justice system despite significant evidence  of shortcomings, such as the linking of criminal datasets to patterns of discriminatory policing \citep{propublica2016, lum2016predict, richardson2019dirty}. Beyond the domain of criminal justice, there are numerous instances of predictive algorithms perpetuating social harms in everyday interactions, including examples of facial recognition systems failing to detect Black faces and perpetuating gender stereotypes \citep{buolamwini2018gender, keyes2018misgendering, stark2019facial}, hate speech detection algorithms identifying Black and queer vernacular as ‘toxic’ \citep{sap2019risk}, new recruitment tools discriminating against women \citep{dastin2018amazon}, automated airport screening-systems systematically flagging trans bodies for security checks \citep{costanza2018design}, and predictive algorithms used to purport that queerness can be identified from facial images alone \citep{arcas2018orientation}.

The current discourse on fairness, accountability and transparency in sociotechnical systems\footnote{For example, see \href{https://facctconference.org/}{ACM Conference on Fairness, Accountability, and Transparency.}}, under which many of these cases are discussed, can be further enriched if these inequities are historically contextualised in global systems of racial capitalism, class inequality, and heteronormative patriarchy, rooted in colonial history \citep{bhattacharyya2018rethinking, sokoloff2008introduction}. In the case of racial capitalism, similarly to \citet{ricaurte2019data}, \citet{couldry2019costs}, and \citet{milan2019big},we propose that institutionalised harms replicated by automated decision-making tools should be understood as continuous to, and inextricably linked with, ``histories of racist expropriation'', and that ``only by tracking the interconnections between changing modes of capitalism and racism that we can hope to address the most urgent challenges of social injustice'' \citep{bhattacharyya2018rethinking}.

A decolonial framework helps connect instances of algorithmic oppression to wider socio-political and cultural contexts, enabling a geographically, historically and intersectionally expansive analysis of risks and opportunities pertaining to AI systems. Notably, it allows for the analysis to move beyond North American or European identity frameworks or definitions of harms. By connecting instances of algorithmic oppression across geographies, new approaches that consider alternative possibilities of using technology in socially complex settings in more critical and considered ways will emerge, and so too will designs that incorporate inclusive and well-adapted mechanisms of oversight and redress from the start. 

\subsection{Algorithmic Exploitation} 

Algorithmic exploitation considers the ways in which institutional actors and industries that surround algorithmic tools take advantage of (often already marginalised) people by unfair or unethical means, for the asymmetrical benefit of these industries. The following examples examine colonial continuities in labour practices and scientific experimentation in the context of algorithmic industries.

\subsubsection*{Site 2: Ghost Workers}

Many of the recent successes in AI are possible only when the large volumes of
data needed are annotated by human experts to expose the common-sense elements that make the data useful for a chosen task. The people who do this labelling for a living, the so
called `ghost workers' \citep{gray2019ghost}, do this work in remote settings,
distributed across the world using online annotation platforms or within
dedicated annotation companies. In extreme cases, the labelling is done by prisoners \citep{haomit2019} and the economically vulnerable \citep{yuannyt2018}, in geographies with limited labour laws. This is a complicated scenario. On one hand such distributed work enables economic development, flexibility in working, and new forms of rehabilitation. On the other, it establishes a form of knowledge and labour extraction, paid at very low rates, and with little consideration for working conditions, support systems and safeties.

A decolonial lens shifts our view towards understanding how colonial history
affects present day labour regulation and enforcement \citep{ronconi2015enforcement}, and how the capacity to mobilise production and outsource services across borders allows
industries to take advantage of present-day post-colonial economic inequalities in order to “reorganize production in ways and places that reduce manufacturing costs and enhance corporate profit" \citep{gomberg2018review, wallerstein1987world}. Logics of colonial extraction and exploitation have ``mutated but also maintain continuity in the present day'', supporting post-colonial economic inequalities which have been empirically demonstrated to be tied to historic colonial activity \citep{bruhn2012good, fanon2007wretched}.  
Data generation and processing presents opportunities for extraction and excavation within data mining industries, which are arguably “ingrained in practices and techniques of extraction [and] is a kind of colonial imprint” \citep{mezzadra2017multiple}, as demonstrated in part by the location of many ghost workers in previously colonised geographies. 

\subsubsection*{Site 3: Beta-testing}
There is a long and well-documented history on the exploitation of marginalised populations for the purpose of scientific and technological progress. Colonies of the British empire ``provided a laboratory for experimenting with new forms of medical and scientific practice'' \citep{senior2018caribbean, tilley2014conclusion}.
There has been a historic continuity of scientific experimentation on African Americans, from early experimentation on black enslaved women and infants in the 19th century that is foundational to the field of gynaecology \citep{washington2006medical}, to the Tuskegee syphilis study \citep{brandt1978racism}. Such experimental practices continue to colour the establishment of socio-economic development schemes in previously colonised countries, often by former colonisers \citep{bonneuil2000development}. 

It is with this historic lens that we examine the practice of beta-testing, which is the testing and fine-tuning of early versions of software systems to help identify issues in their usage in settings with real users and
use-cases. In the testing of predictive systems we find several clearly
exploitative situations, where organisations use countries outside of their own
as testing grounds---specifically because they lack pre-existing safeguards and
regulations around data and its use, or because the mode of testing would
violate laws in their home countries \citep{unctad2013information}. This phenomenon is known as \textit{ethics dumping}: the export of harms and unethical research practices by companies to marginalised and vulnerable populations or to low and middle income countries, and which often aligns ``with the old fault lines of colonialism'' \citep{schroeder2018ethics}.  The counterpoint to ethics dumping is \textit{ethics shirking}: what is not done to protect people when harms emerge beyond what is demanded from legal or regulatory frameworks \citep{floridi2019translating}. 

As an example, Cambridge Analytica (CA) elected to beta-test and develop algorithmic tools for the 2017 Kenyan and 2015 Nigerian elections, with the intention to later deploy these tools in US and UK elections. Kenya and Nigeria were chosen in part due to the weaker data protection laws compared to CA's base of operations in the United Kingdom---a clear example of ethics dumping. These systems were later found to have actively interfered in electoral processes and worked against social cohesion \citep{nyabola2018digital}. A critical decolonial approach would, for example, lead us to ask early on why the transgression of democratic processes by companies such as CA only gained international attention and mobilisation after beginning to affect western democratic nations.

In another case of beta-testing, the deployment of predictive algorithms for child welfare interventions by the New Zealand government initially targeted M\={a}ori, the indigenous people of New Zealand, who have long experienced institutional racism \citep{vaithianathan2013children, gavighan2019government}. Analogously, the data analytics firm Palantir was found to have experimentally deployed predictive algorithms in the city of New Orleans (in concert with the police department) without  public approval. These tools were used to target specific individuals and neighbourhoods for police surveillance, and disproportionately impacted African-Americans \citep{bullingtonnola2018}. These are all cases that cannot be viewed ahistorically, e.g., for African Americans there is a historic continuity from 19th century gynaecology experimentation, to the 20th century Tuskegee experiments, to 21st century predictive policing and beta-testing.  

The perspective of historic continuity provided by decolonial theory raises important questions of accountability, responsibility, contestation, and recourse, which become increasingly necessary in entangled settings of low regulation, combined with deficits of localised expertise and contextualised historic knowledge within firms expanding into new markets. Risks are likely to arise if we neglect to explore the current variation of ethical standards based on identity and geography, as well as how algorithms and automated systems interact with existing social stratification at both local and global levels. 

\subsection{Algorithmic Dispossession} 

Algorithmic dispossession, drawing from  \citet{harvey2004new} and \citet{thatcher2016data},  describes how, in the growing digital economy, certain regulatory policies result in a centralisation of power, assets, or rights in the hands of a minority and the deprivation of power, assets, or rights from a disempowered majority. The following examples examine this process in the context of international AI governance (policy and ethics) standards, and AI for international social development. 

\subsubsection*{Site 4: National Policies and AI Governance}

Power imbalances within the global AI governance discourse encompasses issues of data inequality and data infrastructure sovereignty, but also extends beyond this. We must contend with questions of \textit{who} any AI regulatory norms and standards are protecting, who is empowered to project these norms, and the  risks posed by a minority continuing to benefit from the centralisation of power and capital through mechanisms of dispossession \citep{thatcher2016data, harvey2004new}.  As \citet{jasanoff2018global} remind us, we must be mindful of ``who sits at the table, what questions and concerns are sidelined and what power asymmetries are shaping the terms of debate''. 

A review of the global landscape of AI ethics guidelines \citep{jobin2019global} pointed out the “under-representation of geographic areas such as Africa, South and Central America and Central Asia” in the AI ethics debate. The review observes a power imbalance wherein ``more economically developed countries are shaping this debate more than others, which raises concerns about neglecting local knowledge, cultural pluralism and the demands of global fairness''. A similar dynamic is found when we examine the proliferation of national policies
on AI in countries across the world \citep{dutton2018}. In some views, this is a
manifestation of a new type of geopolitics amongst `AI superpowers'
\citep{lee2018ai}, and a rise of `AI nationalism', where nations wrangle to
spread a preferred view of policy, applied approaches and technical services
\citep{hogarth_nationalisms, edgerton2007contradictions}. We are quickly led to one possible scene of coloniality by \citet{lee2017real}: ``Unless they [developing countries] wish to plunge their people into poverty, they will be forced to negotiate with whichever country supplies most of their AI software—China or the United States—to essentially become that country’s economic dependent''. It can be argued that the agency of developing countries is in these ways undermined, where they “cannot act unilaterally to forge their own rules”and cannot expect prompt protection of their interests \citep{pathways2019}. 

Such concerns were demonstrated at the 2019 G20 summit, where a number of developing countries including India, Indonesia and South Africa refused to sign the Osaka Track, an international declaration on data flows \citep{kanth2019india}, because the interests, concerns and priorities of these countries were not seen to be represented in the document. 
The undermining of interests and agency of developing countries is also a relevant issue vis à vis the OECD AI Principles \citep{oecdAI2019}. As these guidelines are adopted and enforced by partner countries around the world, we see analogous  concerns surfacing around exclusionary path-dependencies and first-mover advantages \citep{pathways2019}. Additionally, AI governance guidelines risk being replicated across jurisdictions in a way that may be incompatible with the needs, goals and constraints of developing countries, despite best efforts \citep{pathways2019}.

There are clear hierarchies of power within these cases of policy development, which can be
analysed using the aforementioned metropole-periphery model. It is metropoles (be it government or industry) who are empowered to impose normative values and standards, and may do so at the ``risk of forestalling alternative visions'' \citep{greene2019better}. A metropole-periphery model draws attention to the need to represent values, interests, concerns and priorities of resource-constrained countries in AI governance processes, as well as the historic dynamics that prevent this. Decolonial theory offers AI policy makers a framework to interrogate imbalances of power in AI policy discourse, understand structural dependencies of developing countries, question ownership of critical data infrastructures, and assess power imbalances in product design/development/deployment of computational technologies \citep{irani2010postcolonial} as well as the unequal distribution of risks and economic benefits. 

\subsubsection*{Site 5: International Social Development}

Much of the current policy discourse surrounding AI in developing countries is in economic and social development where advanced technologies are propounded as solutions for complex developmental scenarios, represented by the growing areas of AI for Good and AI for the Sustainable Development Goals (AI4SDGs) \citep{vinuesa2020role, floridi2018ai4people, tomavsev2020ai}.
In this discourse, \citet{green2019good} proposes that ``good isn’t good enough'', and that there is a need to expand the currently limited and vague definitions within the computer sciences of what `social good' means.

To do so,  we can draw from existing analysis of ICT for Development, which are often based on historical analysis and decolonial critique \citep{irani2010postcolonial, toyama2015geek}. These critiques highlight concerns of dependency, dispossession, or ethics dumping and shirking, as discussed earlier \citep{schroeder2018ethics}.
Such critiques take renewed form as AI is put forward as a needed tool for social development. Where a root cause of failure of developmental projects lies in default attitudes of paternalism, technological solutionism and predatory inclusion, e.g., `surveillance humanitarianism' \citep{latonero2019, vinuesa2020role}, decolonial thinking shifts our view towards systems that instead promote active and engaged political community. This implies a shift towards the design and deployment of AI systems that is driven by the the agency, self-confidence and self-ownership of the communities they work for, e.g, adopting co-development strategies for algorithmic interventions alongside the communities they are deployed in \citep{katell2020toward}. 

Co-development is one potential strategy within a varied toolkit supporting the socio-political, economic, linguistic and cultural relevance of AI systems to different communities, as well as shifting power asymmetries.  
A decolonial view offers us tools with which to engage a reflexive evaluation and continuous examination of issues of cultural encounter, and a drive to question the philosophical basis of development \citep{kiros1992moral}. With a self-reflexive practice, initiatives that seek to use AI technologies for social impact can develop the appropriate safeguards and regulations that avoid further entrenching exploitation and harm, and can conceptualise long-term impacts of algorithmic interventions with historical continuities in mind.

\section{Tactics for a Decolonial AI}
\label{sect:decolonisingAI}

By fusing the fields of artificial intelligence and decolonial theories we can take advantage of historical hindsight to develop new tools of foresight and practice. In so doing, we can establish a decolonial AI
that can re-create the field of artificial intelligence in ways that strengthens its empirical basis, while anticipating and averting algorithmic colonialism and harm. 

The five sites of coloniality in the previous section cast the applications of AI research (its products and predictions---AI as object) and the structures that support it (data, networks and policies---AI as subject) as expressions of the coloniality of power \citep{quijano2000coloniality, quijano2007coloniality, mignolo2007introduction,maldonado2007coloniality,ndlovu2015decoloniality}, and of technological power \citep{ricaurte2019data,couldry2019costs,ali2016brief}.
This leads us to seek the decolonisation of power, whose aim is dismantle harmful power asymmetries and concepts of knowledge, turning us instead towards a ``pluriversal epistemology of the future'' \citep{mignolo2012local} that unlike universalisms, acknowledges and supports a wider radius of socio-political, ecological, cultural, and economic needs. 

In this final section, we aim to develop sets of \textit{tactics} for the future development of AI, which we believe open many areas for further research and action. Tactics do not lead to a conclusive solution or method, but instead to the ``contingent and collaborative construction of other narratives'' \citep{philip2012postcolonial}. Our tactics resonate with the proposals for reforming epistemic practice articulated by many other scholars. \citet{couldry2019costs} put forward a vision for decolonising data relations by exploring six tasks---reframing what data is for, restoring well-being, naming alternative world views, gendering, protecting, and creating new forms of social relations---that must all be oriented towards social goals.  \citet{ricaurte2019data} points to needed change in data governance and regimes, addressing technological sovereignty and agency, addressing the impact of technological systems on ecological systems, and the need to imagine alternative digital futures. \citet{benjamin2019race} asks us to retool solidarity and reimagine justice, by rethinking design, developing coded equity audits, and developing abolitionist tools that reimagine technology.

We submit three tactics for future AI design---supporting a critical technical practice of AI, establishing reciprocal engagements and reverse pedagogies, and the renewal of affective and political community---based on lessons of resistance and recovery from historical and decolonial criticism, and grounded within already-existing work that shows how these tactics might be enacted in practice. 

\subsection{Towards a Critical Technical Practice of AI}
The basis of decolonial AI rests in a self-reflexive approach to developing and deploying AI that recognises power imbalances and its implicit value-systems. It is exactly this type of framework that was developed by \citet{agre1997toward}, who described a shift towards a \textit{Critical Technical Practice of AI} (CTP).
Critical technical practices take a middle ground between the technical work of developing new AI algorithms and the reflexive work of criticism that uncovers hidden assumptions and alternative ways of working. CTP has been widely influential, having found an important place in human-computer interactions (HCI) and design \citep{dourish2004reflective, sengers2006reflective}. By infusing CTP with decoloniality we can place a productive pressure on our technical work, moving beyond good-conscience design and impact assessments that are undertaken as secondary tasks, to a way of working that continuously generates provocative questions and assessments of the politically-situated nature of AI.

The role of \textit{practice} in this view is broad by necessity. Recent research, in both AI and Science and Technology Studies (STS), highlights the limitations of purely technological approaches to addressing the ethical and social externalities of AI. Yet, technical approaches can meaningfully contribute when they appropriately reflect the values and needs of relevant stakeholders and impacted groups \citep{selbst2019fairness}. This context-aware technical development that CTP speaks to---which seeks to consider the interplay between social, cultural, and technical elements---is often referred to as heterogeneous engineering \citep{law1987technology}. As a result, a heterogeneous-critical practice must encompass multiple approaches for action: in research, organising, testing, policy, and activism.
We explore five topics constituting such a practice: algorithmic fairness, AI safety, equity and diversity, policy-making, and AI as a decolonising tool. 

\noindent\textbf{Fairness.} Research in \textit{algorithmic fairness} \citep{nissenbaum2001computer, dwork2012fairness, barocas2016big} has recognised that efforts to generate a fair classifier can still lead to discriminatory or unethical outcomes for marginalised groups, depending on the underlying dynamics of power; because a `true' definition of fairness is often a function of political and social factors. \citet{quijano2000coloniality} again speaks to us, posing questions of who is protected by mainstream notions of fairness, and to understand the exclusion of certain groups as ``continuities and legacies of colonialism embedded in modern structures of power, control, and hegemony''.
Such questions speak to a critical practice whose recent efforts, in response, have proposed fairness metrics that attempt to use causality \citep{chiappa2018causal, mitchell2018prediction, nabi2018fair, madras2019fairness} or interactivity \citep{canetti2019soft, jung2019eliciting} to integrate more contextual awareness of human conceptions of fairness.

\noindent\textbf{Safety.} The area of \textit{technical AI safety} \citep{amodei2016concrete, Raji2020a} is concerned with the design of AI systems that are safe and appropriately align with human values. The philosophical question of value alignment arises, identifying the ways in which the implicit values learnt by AI systems can instead be aligned with those of their human users. A specification problem emerges when there is a mismatch between the ideal specification (what we want an AI system to do) and the revealed specification (what the AI system actually does). This again raises questions that were posed in the opening of whose values and goals are represented, and who is empowered to articulate and embed these values---introducing discussions of utilitarian, Kantian, and volitional views on behaviour, and on the prevention and avoidance of undesirable and unintended consequences \citep{gabriel2020artificial}. Of importance here, is the need to integrate discussions of social safety alongside questions of technical safety.
 
\noindent\textbf{Diversity.} With a critical lens, efforts towards greater equity, diversity and inclusion (EDI) in the fields of science and technology are transformed from the prevailing discourse that focuses on the business case of building more effective teams or as being a moral imperative \citep{rock2016diverse}, into \textit{diversity as a critical practice} through which
issues of homogenisation, power, values, and cultural colonialism are directly
confronted. Such diversity changes the way teams and organisations think at a fundamental
level, allowing for more intersectional approaches to problem-solving to bee taken \citep{d2020data}.

\noindent\textbf{Policy.} There is growing traction in \textit{AI governance} in developing countries to encourage localised AI development, 
such as the initiatives by UNESCO, UN Global Pulse’s AI policy support in Uganda and Ghana \citep{unactivitiesAI2019} and Sierra Leone’s National Innovation and Digital Strategy \citep{dsti2019},
 or in structuring protective mechanisms against exploitative or extractive data practices \citep{gray2019ghost}. Although there are clear benefits to such initiatives, international organisations supporting these efforts are still positioned within metropoles, maintaining the need for self-reflexive practices and considerations of wider political economy \citep{pathways2019}.

\noindent\textbf{Resistance.} The \textit{technologies of resistance}  have often emerged as a consequence of opposition to coloniality, built by self-organising communities to ``bypass dynamics and control of the state and corporations'' \citep{steiner1994technologies, milan2013social}. A renewed critical practice can also ask the question of whether AI can itself be used as a decolonising tool, e.g., by exposing systematic biases and sites of redress. For example, \citet{chen2019can} instantiate this idea of using AI to assess systemic biases in order to reduce disparities in medical care, by studying mortality and 30-day psychiatric readmission with respect to race, gender, and insurance payer type as a proxy for socioeconomic status. 
Furthermore, although AI systems are confined to a specific sociotechnical framing, we believe that they can be used as a decolonising tool while avoiding a techno-solutionism trap. When AI systems can be adapted to locally-specific situations in original ways, they can take a renewed role as `creole technologies' that find positive and distinctive use at scale, and outside their initially-conceived usage \citep{edgerton2007creole}. 

\subsection{Reciprocal Engagements and Reverse Tutelage}

Research in post-colonial studies increasingly highlights the essential
role that colonised peoples themselves, through insurgence, activism and
organisation, had in changing the colonial view in the metropole
\citep{gopal2019insurgent, gandhi2006affective}. Despite colonial power, the historical record shows that colonialism was never only an act of imposition. In a reversal of roles, the metropole often took lessons from the periphery, establishing a reverse tutelage between centre and periphery. A modern critical practice would seek to use this decolonial imperative to develop a double vision: actively
identifying centres and peripheries that make reverse tutelage and the resulting pedagogies of reciprocal exchange part of its
foundations, while also seeking to undo colonial binarisms.

Reverse tutelage directly speaks to the philosophical questions of what
constitutes knowledge. There remains a tension between a view of knowledge as
absolute and of data that, once enough is collected, allows us to form complete and encompassing abstractions of the
world, versus a view of knowledge that is always incomplete and subject to
selections and interpretation under differing value systems. These oppositional
views of knowledge have been explored in different ways,
such as in the important anthropological work by \citet{forsythe1993engineering, forsythe2001studying} on knowledge in AI, in the genealogy of statistics as
`the moral science' \citep{hacking2015biopower}, and through `new data epistemologies' \citep{milan2016alternative}. Deciding what counts as valid
knowledge, what is included within a dataset, and what is ignored and
unquestioned, is a form of power held by AI researchers that cannot be left
unacknowledged. It is in confronting this condition that decolonial science, and
particularly the tactic of reverse tutelage, makes its mark. We put forward three modes---of dialogue, documentation, and design---through which reciprocal tutelage can be enacted.

\noindent\textbf{Dialogue.} Reverse pedagogies create a decolonial shift from paternalistic towards solidaristic modes of working that can be achieved by systems of meaningful \textit{intercultural dialogue}. Such dialogue is core to the field of intercultural digital ethics, which asks questions of how technology can support society and culture, rather than becoming an instrument of cultural oppression and colonialism \citep{capurro2018intercultural}. Intercultural ethics emphasises the limitations and coloniality of universal ethics---often the generalisation of dominant rather than inclusive ethical frameworks---and finds an alternative in pluralism, pluriversal
ethics and local designs \citep{escobar2011sustainability, ess2006ethical}. One approach to reverse pedagogies is invoked by \citet{arora2019decolonizing} in the field of privacy research, by interrogating the empirical basis of privacy studies, and calling for an `epistemic disobedience' and a reliance on shifting roles of the metropole and periphery.

\noindent\textbf{Documentation.} New
frameworks have been developed that make explicit the representations of
knowledge assumed within a data set and within deployed AI-systems. Data sheets
for data sets aim to summarise what is and is not contained within a data set
\citep{gebru2018datasheets}, and similar explicit assessments for AI systems exist using the model cards framework \citep{mitchell2019model}. The example in \citet{mitchell2019model} on toxicity scoring provides a simple and powerful example of reverse pedagogy, wherein affected users exposed the system's  limitations that led to documented improvements in its subsequent releases.

\noindent\textbf{Design.} There is now also a growing understanding of approaches
for meaningful community-engaged research \citep{mikesell2013ethical}, using frameworks like the IEEE Ethically Aligned Design
\citep{ieee2016ethically}, technology policy design frameworks like Diverse
Voices \citep{young2019toward}, and mechanisms for the co-development of algorithmic accountability through participatory action research \citep{katell2020toward}. The framework of citizens' juries have also been
used to gain insight into the general public's understanding of the role and
impact of AI \citep{balaram2018artificial}. 

A critical viewpoint may not have
been the driver of these solutions, and these proposals are themselves subject
to limitations and critique, but through an ongoing process of criticism and
research, they can lead to powerful mechanisms for reverse tutelage in AI design
and deployment.

\subsection{Renewed Affective and Political Communities}

How we build a critical practice of AI depends on the strength of political communities to shape the ways they will use AI, their inclusion and ownership of advanced technologies, and the mechanisms in place to contest, redress and reverse technological interventions.
The systems we described in section \ref{sect:coloniality_sites}, although ostensibly
developed to support human decision-makers and communities, failed to meaningfully engage
with the people who would be the targets of those systems, cutting off these avenues of ownership, inclusion and justice.
The historical record again shows that these situations
manifest through paternalistic thinking and imbalances in authority and choice; produced by the
hierarchical orders of division and binarisation established by coloniality \citep{gopal2019insurgent, said2012culture, fanon2008black, nandy1989intimate}. The decolonial imperative asks for a move from attitudes of technological benevolence and paternalism towards solidarity. This principle enters amongst the core of decolonial tactics and foresight, speaking to the larger goal of decolonising power.

The challenge to solidarity lies in how new types of political community can be created that are able to reform systems of hierarchy, knowledge, technology and culture at play in modern life. 
One tactic lies in embedding the tools of decolonial thought within AI design and research. Contrapuntal analysis \citep{said2012culture} is one important critical tool that actively leads us  to expose the habits and codifications that embed questionable binarisms---of metropole and periphery, of west and the rest, of scientists and humanist, of natural and artificial---in our research and products. 
Another tactic available to us lies in our support of grassroots
organisations and in their ability to create new forms of affective community, elevate intercultural dialogue, and demonstrate the forms of solidarity and alternative community that are already possible. Many such groups already exist, particularly in the field of AI, such as Data for Black Lives \citep{ goyanes2018}, the Deep Learning Indaba \citep{gershgorn2019Africa}, Black in AI and Queer in AI, and are active across the world.

The advantage of historical hindsight means that we can now recover the principles of living that were previously made incompatible with life by colonial binaries. Friendship quickly emerges as a ``lost trope of anticolonial thought'' \citep{gandhi2006affective}. This is a political friendship that has been expanded in many forms: by \citet{el2010open} using the concept of \textit{aimance} , in the politics of friendship \citep{derrida1993politics}, and as \textit{affective communities} \citep{gandhi2006affective} in which developers and users seek alliances and connection outside possessive forms of belonging. Other resources are also available: within the principles of political love \citep{fanon2008black, zembylas2017love, butorac2018hannah}, in moral philosophies that recognise both material and immaterial development \citep{kiros1992moral}, and in philosophies such as Ubuntu \citep{ramose1999african} and
Oneness \citep{wong2012dao}. And the decolonial principles we described---moves
from benevolence to solidarity, enabling systems of reverse tutelage,
harnessing technological solutions, creating a critical technical practice---when combined, shape the creation of these
new types of political and affective community. These ideas are relevant to the work of AI at its most fundamental because, recalling the use-cases in section \ref{sect:coloniality_sites}, AI
is shaped by, and shapes the evolution of contemporary political community. 

Finally, these views of AI taken together lead us quickly towards fundamental
philosophical questions of what it is to be human---how we relate and live with
each other in spaces that are both physical and digital, how we navigate
difference and transcultural ethics, how we reposition the roles of culture
and power at work in daily life---and how the answers to these questions are reflected in the AI systems we build. Here alone is there an ongoing need for research and action, and to which historical hindsight and technological foresight can make significant contributions.

\section{Conclusions}
\label{sect:concl}

This paper aims to provide a perspective on the importance of a critical science approach, and in particular of decolonial thinking, in understanding and shaping the ongoing advances in AI. Despite these fields having operated mostly apart, our hope is that decolonial theories will expand the practitioner's ability, including ourselves, to ask critical questions of technology research and design, to imagine alternative realities, to question who is developing AI and where, and to examine the roles of culture and power embedded in AI systems; reinvigorating Agre's \citeyearpar{agre1997toward} vision of a critical technical practice of AI.

Operationalising this critical practice will require not only foresight and case study research, but also approaches that support new research cultures, along with innovative technical research in domains such as fairness, value alignment, privacy, and interpretability. Moreover, there is a strong need to develop new methodologies for inclusive dialogue between stakeholders in AI development, particularly those in which marginalised groups have meaningful avenues to influence the decision-making process, avoiding the potential for predatory inclusion, and continued algorithmic oppression, exploitation, and dispossession.

Any commitment to building the responsible and beneficial AI of the future ties us to the hierarchies, philosophy and technology inherited from the past, and a renewed responsibility to the technology of the present. To critically engage with that inheritance, to avert algorithmic coloniality, to reveal the relations of power that underlie technology  deployment, to recover the principles of living that were previously made incompatible with life, and to create new forms of political and affective community, are the tasks for which decolonial theories will provide key resources---whether these be through heterogeneous practice, intercultural dialogue, creole AIs, reverse tutelage, or a politics of friendship and love.
The years ahead will usher in a wave of new scientific breakthroughs and technologies driven by AI, making it incumbent upon AI communities to strengthen the social contract through ethical foresight and the multiplicity of intellectual perspectives available to us, aligned with the goal of promoting beneficence and justice for all.

\section*{Acknowledgements}
\vspace{-4mm}
We thank our reviewers and editor whose positive feedback and mentorship has improved this paper. We are grateful to Desha Osborne, Irene Solaiman, Momin Malik, and Vafa Ghazavi, who have influenced our thinking. We are also grateful to many of our colleagues for their support, including Ben Coppin, Brittany Smith, Courtney Biles, Dorothy Chou, Heiko Strathmann, Iason Gabriel, James Besley, Kelly Clancy, Koray Kavukcuoglu, Laura Weidinger, Murray Shanahan, and Sean Legassick. 

\vspace{4mm}
\noindent Disclaimer:
Any opinions presented in this paper represent the personal views of the authors and do not necessarily reflect the official policies or positions of their organisations.

\bibliographystyle{spbasic}
\bibliography{refs}
\end{document}